\documentstyle[aps,epsfig,prl]{revtex}
\begin{document}
\title{Principals of the theory of light reflection and absorption
by low-dimensional semiconductor objects in quantizing magnetic
fields at monochromatic and pulse excitations}
\author{I. G. Lang, L. I. Korovin}
\address{A. F. Ioffe Physical-Technical Institute, Russian
Academy of Sciences, 194021 St. Petersburg, Russia}
\author {S. T. Pavlov\dag\ddag}
\address {\dag Facultad de Fisica de la UAZ, Apartado Postal C-580,
98060 Zacatecas, Zac., Mexico \\
\ddag P. N. Lebedev Physical Institute, Russian Academy of
sciences, 119991 Moscow, Russia}
\twocolumn[\hsize\textwidth\columnwidth\hsize\csname
@twocolumnfalse\endcsname
\date{\today}
\maketitle \widetext
\begin{abstract}
\parbox{6in} {The bases of the theory of light reflection and absorption
by low-dimensional semiconductor objects (quantum wells, wires and
dots) at both monochromatic and pulse irradiations and at any form
of light pulses are developed. The semiconductor object may be
placed in a stationary quantizing magnetic field.

 As an example the case of normal light incidence on a quantum well (QW) surface is
 considered. The width of the QW may be comparable to the light wave length
 and number of energy levels of electronic excitations is arbitrary.
 For Fourier-components of electric fields the integral equation (similar to the Dyson-equation)
 and solutions of this equation for some individual cases are
 obtained.}
\end{abstract}

] \narrowtext

  The interest to optics time-dependent effects in semiconductor objects is great in last
years [1-4]. This is connected with successful engineering of
short light pulses what allows to investigate coherent phenomena
in processes of light interactions with elementary excitations in
various systems.

 If a semiconductor object is irradiated with a short light pulse there appears
a pulse of a secondary light radiation the form of which
 differs essentially from the form of primary
 light pulse and  bears also the information on exited states
 in the object, for example about
 life-times of electron-hole pairs (EHP), about
 splittings of magneto-polaron energy spectrum and so on.

 In general the secondary light radiation from material objects is powerful
 means of research of their internal structure.
 Both at monochromatic and at a pulse light irradiation there are two sorts
 secondary light radiation. For example, at irradiation of low-dimensional
  semiconductor objects
  the secondary light radiation of the first sort causes light
 reflection from these objects which may be resonant if
 the frequency $ \omega_l $ of stimulating light coincides with the frequency $ \omega_0 $
 of one of discrete energy levels of an electronic system. In bulk
  semiconductors the secondary radiation of the first sort causes difference
 of true electromagnetic fields from stimulating fields, i.e. causes
 a deviation of the dielectric susceptibility $ \varepsilon $ from unit.

 The second sort of secondary light radiation is light scattering, for example Raman
 scattering, which can not be described in terms of
 the dielectric susceptibility.

 How does any secondary radiation from material objects appear?
 Stimulating light creates in systems of charged particles induced alternating
 electric currents and charge density fluctuations. Current and
 charge fluctuations
 cause secondary electromagnetic fields. This chem is applicable equally to
 bulk bodies and to low-dimensional systems, for instance, to quantum
 wells.

 If to average densities of induced currents and charges (on the ground state of system,
 for example, in a case of zero temperature) and
 to calculate the induced electromagnetic fields, we obtain the secondary light radiation
 of the first sort. The secondary light radiation of the second sort, i.e. light scattering,
 is caused by fluctuations of induced current and charge densities.
 Below we investigate only the secondary light radiation of the first
 sort and light absorption.

 Modern semiconductor technologies allow to make high quality quantum
 wells, when radiative broadening of an absorption line may
 be comparable to the contributions of non-radiative relaxation mechanisms or to exceed
 them. In such situation it is impossible to be limited by the lowest approximation
  on interaction of electrons
 with electromagnetic fields and it is necessary to take into account all orders
  of this interaction [5-13].

 Below principals of the theory of the secondary radiation of the
 first sort from low-dimensional semiconductor objects are developed. The accent is made on
 situation when the object is placed in a quantizing stationary magnetic field.
 Results are applied to a case of light pulse irradiation with any pulse
 forms.

 The statement is constructed as follows. In section I the expressions for
 averaged values of current and charge densities induced by a weak electromagnetic
 field in a system of limited in space charged particles  are given. These expressions
 are applicable in case of any stationary potential, any interaction
 between particles and any constant magnetic fields.
 Contributions containing electric fields and derivative of electrical fields
 on coordinates are separated. In further last contributions are considered small and
 are not taken into account.

 Further in sections II-VII the averaged density of induced
 currents in low-dimensional
 semiconductor objects is calculated without taking into account the Coulomb
 interaction between electrons and holes and in section VII
 the result is generalized with taking into account of the excitonic effect.

 In section VIII the concept of the conductivity tensor
 $ \sigma _ {\alpha\beta} ({\bf k}, \omega | \bf r) $ depending on
 spatial coordinates due to the spatial heterogeneity of low-dimensional
 semiconductor objects is introduced. The general formula
 for the tensor $ \sigma _ {\alpha\beta} ({\bf k}, \omega | \bf r) $ is applicable
 for quantum wells, wires and dots.

 In section IX  the conductivity tensor for a case of quantum wells in
 zero and quantizing magnetic fields is calculated. In section X the averaged density
of induced currents is calculated for a specific case of normal
light incidence on a
 quantum well surface.

 In section XI the model appropriate to two degenerated valence bands
 and simplifying expressions for the averaged density of induced currents is described.
 It is shown that to this model the zero density of induced charges corresponds.

Using the formula for the retarded potential in section XII we
express the vector potential through an integral containing the
averaged density of induced currents. Knowing the vector potential
we calculate induced electric fields. Since the induced current
density  depends on an electric field we obtain some integral
equation for it.

In section XIII the integral equation is transformed with
reference to an approximation of an infinitely deep quantum well.

In sections XIV-XVI the integral equation for electric fields is
solved for special cases. In section XIV the case of many energy
levels of electronic excitations is considered in a narrow
quantum well, the width of which is much less than the length of
a stimulating light wave. In section XV the equation is solved in
a case of many energy levels in a wide quantum well in the lowest
order on the interaction of electromagnetic fields with
electrons. In section XVI electric fields  are precisely
determined in case of one energy level of an excitation in a wide
quantum well.

At last, in section XVII it is shown how the expressions for
induced fields are connected with the form of a stimulating light
pulse.

\section{The exact formulas for the average induced current and charge densities.}
In [14] it is shown that averaged current and charge densities
induced by external weak electromagnetic fields may be expressed
through values of electric fields and their derivatives on
coordinates as follows
\begin{equation}
\label{1}\langle0|j_{1\alpha}({\bf r},
t)|0\rangle=\langle0|j_{1\alpha}({\bf r},
t)|0\rangle_I+\langle0|j_{1\alpha}({\bf r}, t)|0\rangle_{II},
\end{equation}
\begin{equation}
\label{2}\langle0|\rho_{1}({\bf r},
t)|0\rangle=\langle0|\rho_{1}({\bf r},
t)|0\rangle_I+\langle0|\rho_{1}({\bf r}, t)|0\rangle_{II},
\end{equation}
where the subscript "1" means linear approximation on fields,
indexes $ I $ and $ II $ - contributions containing fields and
their derivatives on coordinates, respectively. The results are
obtained
\begin{eqnarray}
\label{3}\langle0|j_{1\alpha}({\bf r},
t)|0\rangle_I&=&{i\over\hbar}\int d{\bf r}^\prime\int_{-\infty}^tdt^\prime\nonumber\\
&\times&\langle0|[j_{\alpha}({\bf r}, t), \bar{d}_{\beta}({\bf
r}^\prime, t^\prime)]|0\rangle E_{\beta}({\bf r}^\prime,
t^\prime),
\end{eqnarray}
\begin{eqnarray}
\label{4}\langle0|j_{1\alpha}({\bf r}, t)|0\rangle_{II}={e\over m
c}\langle0|\bar{d}_{\beta}({\bf r})|0\rangle{\partial
a_{\beta}({\bf r}, t)\over\partial r_\alpha}\nonumber\\
-{i\over\hbar c}\int d{\bf r}^\prime \langle0|[j_{\alpha}({\bf
r}, t)\bar{Y}_{\beta\gamma}({\bf r}^\prime, t^\prime)]|0\rangle
{\partial a_{\beta}({\bf r}^\prime, t^\prime)\over\partial
r_\gamma^\prime},
\end{eqnarray}
\begin{eqnarray}
\label{5}\langle0|\rho_{1}({\bf r}, t)|0\rangle_I
&=&{i\over\hbar}\int d{\bf r}^\prime\int_{-\infty}^tdt^\prime\nonumber\\
&\times&\langle0|[\rho({\bf r}, t), \bar{d}_{\beta}({\bf
r}^\prime, t^\prime)]|0\rangle E_{\beta}({\bf r}^\prime,
t^\prime),
\end{eqnarray}
\begin{eqnarray}
\label{6}\langle0|\rho_{1}({\bf r}, t)|0\rangle_{II}=- {i\over
\hbar c}\int d{\bf
r}^\prime\int_{-\infty}^tdt^\prime\nonumber\\
\times\langle0|\rho({\bf r}, t), \bar{Y}_{\beta\gamma}
|0\rangle{\partial a_{\beta}({\bf r}^\prime,
t^\prime)\over\partial r_\gamma^\prime}.
\end{eqnarray}

The following designations are used: $ \langle0 |... | 0\rangle $
is averaging on the ground state of system, $ {\bf j} ({\bf r}, t)
$ and $ \rho ({\bf r}, t) $ are the operators of current and
charge densities in the  interaction representation
\begin{eqnarray}
\label{7}{\bf j}({\bf r}, t)=e^{i{\cal H}t/\hbar}{\bf j}({\bf
r})e^{-i{\cal H}t/\hbar},\nonumber\\
\rho({\bf r}, t)=e^{i{\cal H}t/\hbar}\rho({\bf r})({\bf
r})e^{-i{\cal H}t/\hbar},
\end{eqnarray}
where $ {\cal H} $ is the Hamiltonian of the system
\begin{eqnarray}
\label{8}{\cal H}={1\over 2m}\sum_iP_i^2+V({\bf r}_1, ... {\bf r}_N),\nonumber\\
{\bf p}_i={\bf P}_i-{e\over c}A({\bf r}_i),~~{\bf
P}_i=-i\hbar{\partial\over\partial{\bf r}_i},
\end{eqnarray}
$ A ({\bf r}) $ is the vector potential appropriating to a
constant quantizing  magnetic field
$$ {\bf H} _ c = {\bf \nabla} \times A ({\bf r}), $$  $ V ({\bf r}_1...
{ \bf r} _ N) $ is the potential energy including an interaction
between particles and external potential. The Hamiltonian (8)
describes the system of $ N $ particles with a charge $ e $ and
mass $ m $.

 The operators $ {\bf j} ({\bf
r}) $ and $ \rho ({\bf r}) $ are determined as
$${\bf j}({\bf
r})=\sum_i{\bf j}_i({\bf r}),~~~\rho({\bf r})=\sum_i\rho_i({\bf
r}),$$
$${\bf j}_i({\bf r})={e\over 2}\{\delta({\bf r}-{\bf r}_i){\bf v}_i+
{\bf v}_i\delta({\bf r}-{\bf r}_i)\},~~{\bf v}_i={{\bf p}_i\over
m},$$
$$\rho_i({\bf
r})=e\delta({\bf r}-{\bf r}_i),~~~~(8a)$$
$ {\bf r} _ i $ is the coordinate of i-th particle. Let us
introduce designations
$$\bar{\bf d}({\bf r})=\sum_i\bar{\bf r}_i\rho_i({\bf r}),$$
$$\bar{Y}_{\beta\gamma}({\bf r})={1\over2}\sum_i(j_{i\gamma}\bar {r}_{i\beta}+
\bar {r}_{i\beta}j_{i\gamma})$$
$$\bar{\bf r}_i={\bf r}_i-\langle0|{\bf r}_i|0\rangle,~~~~~~~~~~~~~~~~~~~(8b)$$
and also
$${\bf a}({\bf r}, t)=-c\int_0^\infty dt^\prime {\bf E}({\bf r}, t^\prime).$$
Fields are considered as classical, the temperature is equal zero.
At a derivation of (1) - (6) we assumed that on the infinitely
removed distances there are no charges and currents, and also that
on times $ t\rightarrow -\infty $ the fields $ {\bf E} ({\bf r},
t) $ and $ {\bf H} ({\bf r}, t) $ are equal 0 what corresponds to
adiabatic switching on of fields.
\section{The  secondary quantization representation.}
Below we do not take into account the contributions with the index
$ II $ containing derivatives from electric fields on coordinates,
considering these contributions as small in comparison to basic
contributions with the index $ I $ The discussion of this question
see in [15]).

Let us pass to the  secondary quantization representation, using
a set of ortho-normalized  wave functions of particles $ \Psi _ m
$ satisfying to conditions
\begin{equation}
\label{9}\int d{\bf r}\Psi_{m^\prime}^*({\bf r})\Psi_m({\bf
r})=\delta_{mm^\prime}.
\end{equation}
Then operators of current and charge densities determined in (8a)
look like
\begin{eqnarray}
\label{10}j_\alpha({\bf r})&=&{e\over 2m}\sum_{m,
m^\prime}\{\Psi_{m^\prime}^*({\bf r})p_\alpha\Psi_m({\bf
r})\nonumber\\&-&\Psi_{m}({\bf r})p_\alpha\Psi_{m^\prime}^*({\bf
r})\}a^+_{m^\prime}a_m,
\end{eqnarray}
\begin{equation}
\label{11}\rho({\bf r})=e\sum_{m, m^\prime}\Psi_{m^\prime}^*({\bf
r})\Psi_m({\bf r})a^+_{m^\prime}a_m.
\end{equation}
The operator $ \bar {\bf d} ({\bf r}) $ determined in (8b) in the
secondary quantization representation looks like
\begin{equation}
\label{12}\bar{d}_\alpha({\bf r})=e(r_\alpha-r_{0\alpha}) \sum_{m,
m^\prime}\Psi_{m^\prime}^*({\bf r})\Psi_m({\bf
r})a^+_{m^\prime}a_m,
\end{equation}
where
\begin{equation}
\label{13} r_{0\alpha}={1\over N}\sum_{m_0 }\int d^3r
\Psi_{m}^*({\bf r})r_\alpha\Psi_m({\bf r}),
\end{equation}
$ m _ 0 $ is a set of $ N $ states occupied  by particles in the
ground state $ | 0\rangle $. Let us notice that  $ r _ \alpha-r _
{0\alpha} $, included in the right hand part of (12), is
invariant relatively a shift of origin of coordinates, i.e. a
replacement $ {\bf r} $ on $ {\bf r} + {\bf R} $ where $ {\bf R}
$ is any vector.

\section{Consideration of semiconductor objects.}
Let us consider a semiconductor quantum well, wire or dot,
containing valence bands and a conductivity band. From group of
indexes $ m $ we allocate indexes $ v $ and $ c $ of valence and
conductivity bands, respectively. We neglect transitions in
higher bands.  We designate other indexes by $ \zeta $.

  Let us calculate the averaged induced current density using initial
  expression (3). Into the RHS of (3)  only non-diagonal
  matrix elements of the operators
  $ j _ \alpha ({\bf r}) $ and $ \bar {d} _ \beta ({\bf r}) $ introduce,
  because
  the operators stand inside of a commutator. Therefore in the RHS (3) we have to substitute
\begin{eqnarray}
\label{14}j_\alpha^{nd}({\bf r})&=&{e\over 2m}\sum_{v, \zeta,
\zeta \prime}\{\Psi_{c\zeta^\prime}^*({\bf
r})p_\alpha\Psi_{v\zeta}({\bf r})\nonumber\\&-&\Psi_{v\zeta}({\bf
r})p_\alpha\Psi_{c\zeta^\prime}^*({\bf
r})\}a^+_{c\zeta^\prime}a_{v\zeta}+h.c.,
\end{eqnarray}
\begin{eqnarray}
\label{15}\bar{d}_\alpha^{nd}({\bf r})=e \sum_{v, \zeta, \zeta
\prime}\Psi_{c\zeta^\prime}^*({\bf r})( r_\alpha-
r_{0\alpha})\Psi_{v\zeta}({\bf
r})\nonumber\\
\times a^+_{c\zeta^\prime}a_{v\zeta}+h.c..
\end{eqnarray}
In (14) and below we consider electrons, therefore $ e = - | e |
$, the mass $ m = m _ 0 $, where $ m _ 0 $ is the free electron
mass. The superscript $ nd $ means a part of the operator having
only non-diagonal matrix elements.

\section{The effective mass approximation.}
Let us consider that object sizes - width $ d $ of a QW or sizes
of wires or dots - are much greater than the lattice constant $ a
$ and the distances on which varies a slowly varying part of the
wave function, is much greater than $ a $. Then the effective
mass approximation is applicable according to which
\begin{equation}
\label{16} \Psi_{\mu\zeta}({\bf r})=u_{0\mu}({\bf
r})\psi_{\mu\zeta}({\bf r}),
\end{equation}
where $ u _ {0\mu} ({\bf r}) $ is the quickly varying
dimensionless factor, $ \psi _ {\mu\zeta} ({\bf r}) $ is the
slowly varying factor, $ \mu = c $ or $ \mu = v $.

In (14) we neglect an action of the operators $ P_\alpha $ on the
slowly varying factor in wave functions (16). Then
 we have approximately
\begin{eqnarray}
\label{17}j_\alpha^{nd}({\bf r})&\simeq&{e\over 2m_0}\sum_{v
}\{u_{0c}^*({\bf r})p_\alpha u_{0v}({\bf r})- u_{0v}({\bf
r})p_\alpha u_{0c}^*({\bf r})\}\nonumber\\
&\times&\sum_{\zeta, \zeta^\prime}\psi_{c\zeta^\prime}^*({\bf r})
\psi_{v\zeta}({\bf r})a^+_{c\zeta^\prime}a_{v\zeta}+h.c.,
\end{eqnarray}
\begin{eqnarray}
\label{18}\bar{d}_\alpha^{nd}({\bf r})&\simeq&e
\sum_{v}u_{0c}^*({\bf
r})u_{0v}({\bf r})(r_\alpha-r_{\alpha 0})\nonumber\\
&\times&\sum_{\zeta, \zeta^\prime}\psi_{c\zeta^\prime}^*({\bf
r})\psi_{v\zeta}({\bf r})a^+_{c\zeta^\prime}a_{v\zeta}+h.c. .
\end{eqnarray}
We get rid of quickly varying factors in the RHS of (17) and
(18). For this purpose we introduce the Fourier components
\begin{eqnarray}
\label{19}j_\alpha({\bf \kappa})=\int d^3r j_\alpha({\bf r})
e^{-i{\bf \kappa}{\bf r
}},\nonumber\\
\bar{d}_\alpha({\bf \kappa})=\int d^3r \bar{d}_\alpha({\bf
r})e^{-i{\bf \kappa}{\bf r }}.
\end{eqnarray}
If $ \kappa a << 1 $  we obtain approximately
\begin{eqnarray}
\label{20}{\bf j}^{nd}({\bf \kappa})&=&{e\over m_0}\sum_v {\bf p
}_{cv}\sum_{\zeta, \zeta \prime}\left\{\int d^3r\psi_{c,
\zeta^\prime}^*({\bf r})\psi_{v\zeta}m({\bf r})e^{-i{\bf
\kappa}{\bf
r}}\right\}\nonumber\\&\times&a^+_{c\zeta^\prime}a_{v\zeta}+h.c.,
\end{eqnarray}
\begin{eqnarray}
\label{21}\bar{{\bf d}}^{nd}({\bf \kappa})&=&\sum_{v}{\bf d}_{cv}
\sum_{\zeta, \zeta \prime}\left\{\int d^3r\psi_{c,
\zeta^\prime}^*({\bf r})\psi_{v\zeta}({\bf r})e^{-i{\bf
\kappa}{\bf
r}}\right\}\nonumber\\&\times&a^+_{c\zeta^\prime}a_{v\zeta}+h.c.,
\end{eqnarray}
where
\begin{eqnarray}
\label{22}{\bf p}_{cv}={1\over\Omega}\int_\Omega d^3r
u_{0c}^*({\bf r}){\bf P}u_{0v}({\bf r}),\nonumber\\
{\bf d}_{cv}={e\over\Omega}\int_\Omega d^3r u_{0c}^*({\bf r}){\bf
r}u_{0v}({\bf r}),
\end{eqnarray}
$ \Omega $ is the volume of an elementary crystal cell  on which
the integration is made.

In the first equality of (22) we have replaced the operator $ {\bf
p} $ (see (8)) on $ {\bf P} $. If the system is in a quantizing
magnetic field $ {\bf H}_c = rot {\cal {\bf A}} ({\bf r}) $ and in
the operator $ {\bf p} $ the term $ - (e/c) {\cal {\bf A}} ({\bf
r}) $ is present, it brings in the small contribution in $ {\bf
p} _ {cv} $ and may be rejected in the effective mass
approximation.

Assuming that in the future we interest only by long wave
components of $ {\bf j} ^ {nd} ({\bf r}) $ and $ \bar {{\bf D}} ^
{nd} ({\bf r}) $, we pass back from $ \kappa $ - representation to
$ r $ - representation and obtain
\begin{eqnarray}
\label{23}{\bf j}^{nd}({\bf r})&=&{e\over m_0}\sum_v {\bf p
}_{cv}\sum_{\zeta, \zeta \prime}\psi_{c\zeta^\prime}^*({\bf
r})\psi_{v\zeta}({\bf
r})\nonumber\\&\times&a^+_{c\zeta^\prime}a_{v\zeta}+h.c.,
\end{eqnarray}
\begin{eqnarray}
\label{24}\bar{{\bf d}}^{nd}({\bf r})&=& \sum_{v}{\bf d}_{cv}
\sum_{\zeta, \zeta \prime}\psi_{c\zeta^\prime}^*({\bf
r})\psi_{v\zeta}({\bf
r})\nonumber\\&\times&a^+_{c\zeta^\prime}a_{v\zeta}+h.c.,
\end{eqnarray}
\section{Electron wave functions  in QW.}
Let us consider two concrete examples of electron long wave
functions in the effective mass approximation in QWs. In the free
electron case
\begin{equation}
\label{25}\psi_{{\bf k}_\perp, l}({\bf
r})={1\over\sqrt{S_0}}e^{i{\bf k}_\perp{\bf r}_\perp}\varphi_l(z),
\end{equation}
where $ S _ 0 $ is the normalization area, the axis $ z $ is
directed perpendicularly to the QW plane, the real function $
\varphi _ l (z) $ corresponds to levels $ l = 1, 2... $ of the
size quantization of electrons. For QWs of a finite depth the
function $ \varphi _ l (z) $ and energy levels, appropriate to
them, are determined, for example, in [16].

The second example is the case of electrons in a QW in quantizing
magnetic field $ {\bf H} _ c $ perpendicular to the QW plane. The
axis $ z $ is directed along a magnetic field. Let us choose the
following gauge of the vector potential
\begin{equation}
\label{26} {\cal {\bf A}} ({\bf r}) = {\cal {\bf A}} (0, xH, 0).
\end{equation}
Then electron wave functions look like
\begin{equation}
\label{27}\psi_{n, k_y, l}({\bf
r})=\Phi_n(x+a_H^2k_y){1\over\sqrt{L_y}}e^{ik_yy}\varphi_l(z),
\end{equation}
\begin{equation}
\label{28} \Phi_n(x)={1\over\sqrt{\pi^{1/2}2^nn!a_H}}H_n(x/a_H)e^
{-x^2/2a_H^2},
\end{equation}
\begin{equation}
\label{29} a_H=\sqrt{c\hbar\over|e|H},
\end{equation}
$ H _ n (t) $ is an Hermitian polynomial, $ L _ y $ is the
normalization length.

\section{The concept of a hole in a valence band.}
Let us consider that quasi-momentum components of a hole $ {\bf k}
_ {h\perp} = - {\bf k} _ {\perp} $ and $ k _ {hy} = -k _ y $ (in
a quantizing magnetic field $ H _ c $) and the operator $ a _
{v\zeta} $ of an electron annihilation in the valence band
 is equal to the operator of a hole creation.
Let us introduce a set of indexes $ \eta $ describing quantum
numbers of an electron-hole pair (EHP) and operator $ a _ \eta ^
+ (a _ \eta) $ of a creation (annihilation) of an EHP. Then from
(23) and (24) we obtain
\begin{equation}
\label{30}{\bf j}^{nd}({\bf r})={e\over m_0}\sum_\eta\{{\bf
p}_{cv}F_\eta^*({\bf r})a^+_\eta+{\bf p}_{cv}^*F_\eta({\bf
r})a_\eta\},
\end{equation}
\begin{equation}
\label{31}\bar{{\bf d}}^{nd}({\bf r})=\sum_\eta\{{\bf
d}_{cv}F_\eta^*({\bf r})a^+_\eta+{\bf d}_{cv}^*F_\eta({\bf
r})a_\eta\},
\end{equation}
where $ F _ \eta ({\bf r}) $ is the EHP wave function at $ {\bf
r} _ e = {\bf r} _ h = {\bf r} $, $ {\bf r} _ e ({\bf r} _ h) $ is
the electron (hole) radius - vector.

In the case of free EHPs in QWs
\begin{equation}
\label{32}F_\eta({\bf r})={1\over S_0}e^{i({\bf k}_{e\perp}+{\bf
k}_{v\perp}){\bf r}_\perp}\varphi^e_{l_e}(z)\varphi^v_{l_v}(z),
\end{equation}
where the set $ \eta $ includes indexes $ v, {\bf k} _ {e\perp},
{\bf k} _ {v\perp}, l _ e, l _ v $. The EHP energy counted from
the ground state energy is equal
\begin{eqnarray}
\label{33}E_\eta=\hbar\omega_\eta=\hbar\omega_g+\varepsilon^e_{l_e}+\varepsilon^v_{l_v}+
{\hbar^2k_{e\perp}^2\over
2m_e}\nonumber\\
+{\hbar^2k_{v\perp}^2\over 2m_v},
\end{eqnarray}
where $ \hbar\omega _ g $ is the band gap, $ m _ e (m _ v) $ is
the electron (hole) effective mass.

In the case of EHPs in QWs in a quantizing magnetic field we have
\begin{eqnarray}
\label{34}F_\eta({\bf
r})=\Phi_{n_e}(x+a_H^2k_{ey})\Phi_{n_v}(x-a_H^2k_{vy})\nonumber\\
\times{1\over
L_y}e^{i(k_{ey}+k_{vy})y}\varphi^e_{le}(z)\varphi^v_{lv}(z),
\end{eqnarray}
the set $ \eta $ includes indexes $ v, n _ e, n _ v, k _ {ey}, k _
{vy}, l _ e, l _ v $. The appropriate energy is equal
\begin{eqnarray}
\label{35}E_\eta=\hbar\omega_g+\varepsilon^e_{l_e}+\varepsilon^v_{l_v}+
\hbar\Omega_{eH}(n_e+1/2)\nonumber\\+\hbar\Omega_{vH}(n_v+1/2),
\end{eqnarray}
\begin{equation}
\label{36}\Omega_{e(v)H}={|e|H_c\over m_{e(v)}c}
\end{equation}
is the cyclotron frequency of electron (hole in a band $ v $). It
is possible to show that the formulas (30) and (31) are
applicable and in those cases when it is essential the Coulomb
interaction between electrons and holes. For example, at $ H _ c
= 0 $ discrete energy levels correspond to excitonic states in QW.
Then $ F _ \eta ({\bf r}) $ is the excitonic wave function at $
{\bf r} _ e = {\bf r} _ h = {\bf r} $, and $ \eta $ is the set of
indexes describing an exciton. In quantizing magnetic fields
Coulomb forces can change a position of energy levels and affect
on the function $ F _ \eta ({\bf r})$. (Conditions of weak
influence of Coulomb
 forces in quantizing magnetic fields see in [27].)

\section{Averaged induced current densities in semiconductor objects.}
It is easy to find a connection between matrix elements $ {\bf p}
_ {cv} $ and $ {\bf
 d} _ {cv} $ determined in (22), if
to use a ratio
$$v_\alpha={i\over\hbar}[{\cal H}, r_\alpha],$$
from which follows
\begin{equation}
\label{37}{\bf d}_{cv}=-{ie\over m_0\omega_g}{\bf p}_{cv}.
\end{equation}
Substituting (37) in (31) and then (30) and (31) in (3), we obtain
\begin{eqnarray}
\label{38}\langle0| j_{1\alpha}({\bf r},
t)|0\rangle={e^2\over\hbar\omega_gm_0}\int
d^3r^\prime\int_{-\infty}^\infty dt^\prime\Theta(t-t^\prime)\nonumber\\
\sum_{\eta,
\eta^\prime}\{p_{cv^\prime\alpha}^*p_{cv\beta}F_{\eta^\prime}({\bf
r})F_{\eta}^*({\bf r}^\prime)\langle0|
a_{\eta^\prime}(t)a^+_\eta(t^\prime)|0\rangle\nonumber\\
+p_{cv^\prime\alpha}p_{cv\beta}^*F_{\eta^\prime}^*({\bf
r})F_{\eta}({\bf r}^\prime)\langle0|
a_{\eta}(t^\prime)a^+_{\eta^\prime}(t)|0\rangle E_\beta({\bf
r}^\prime, t^\prime),
\end{eqnarray}
where $ \Theta (\tau) = 1 $ at $ \tau > 0 $ and $ \Theta (\tau) =
0 $ at $ \tau < 0 $.

Averaging on the ground state gives the result [17, page 48]
\begin{equation}
\label{39}\langle0| a_{\eta^\prime}(t)a^+_\eta(t^\prime)|0\rangle=
\delta_{\eta,
\eta^\prime}e^{i\omega_\eta(t^\prime-t)-(\gamma_\eta/2)|t-t^\prime|},
\end{equation}
where $ \gamma _ \eta $ is the non-radiative broadening of a state
 $ \eta. $

Substituting (39) in (38) and making replacement $ t ^
\prime\rightarrow t + t ^ \prime $, we obtain
\begin{eqnarray}
\label{40}&\langle&0| j_{1\alpha}({\bf r},
t)|0\rangle={e^2\over\hbar\omega_gm_0^2}\sum_{\eta}\nonumber\\
&\times&\left\{p_{cv\alpha}^*p_{cv\beta}F_{\eta}({\bf r}) \int
d^3r\prime F_{\eta}^*({\bf r}^\prime)\int_{-\infty}^0 dt^\prime
e^{i\omega_\eta t^\prime+(\gamma_\eta/2)t^\prime}\right.\nonumber\\
&+&\left.p_{cv\alpha}p_{cv\beta}^*F_{\eta}^*({\bf r}) \int
d^3r\prime F_{\eta}({\bf r}^\prime)\int_{-\infty}^0 dt^\prime
e^{-i\omega_\eta t^\prime+(\gamma_\eta/2)t^\prime}\right\}\nonumber\\
&\times&E_\beta({\bf r}^\prime, t+t^\prime).
\end{eqnarray}
The result (40) is applicable in a wide area, for example, in
case of exciton states at a zero magnetic field and in quantizing
magnetic fields, i.e. at the account of Coulomb interaction of
electrons and holes in those cases when it is essential.
Certainly, the functions $ F _ {\eta} ({\bf r}) $ at the account
 of Coulomb forces will differ from (32) and (34). Besides the
formula (40) is applicable in case of other low-dimensional
semiconductor objects, for example, quantum wires or dots.

\section{Introduction of a conductivity tensor.}
 We write down the expression (40) as
\begin{eqnarray}
\label{41}\langle0| j_{1\alpha}({\bf r}, t)|0\rangle= \int
d^3r\prime \int_{-\infty}^\infty
dt^\prime\sigma_{\alpha\beta}({\bf r}^\prime, t^\prime|{\bf r},
t)\nonumber\\
\times E_\beta ({\bf r}-{\bf r}^\prime, t-t^\prime),
\end{eqnarray}
where $ \sigma _ {\alpha\beta} ({\bf r} ^ \prime, t ^ \prime |
{\bf r}, t) $ is the conductivity tensor (the designation is
borrowed from [18]). It follows from (40)
\begin{eqnarray}
\label{42}&\sigma&_{\alpha\beta}({\bf r}^\prime, t^\prime|{\bf r},
t)={e^2\over\hbar\omega_gm_0^2}
\sum_{\eta}\nonumber\\
&\times&\{p_{cv\alpha}^*p_{cv\beta}F_{\eta}({\bf r})
F_{\eta}^*({\bf r}-{\bf r}^\prime)\Theta(t^\prime)
e^{-i\omega_\eta t^\prime-(\gamma_\eta/2)t^\prime}\nonumber\\
&+&p_{cv\alpha}p_{cv\beta}^*F_{\eta}^*({\bf r})F_{\eta}({\bf
r}-{\bf r}^\prime) \Theta(t^\prime) e^{i\omega_\eta
t^\prime-(\gamma_\eta/2)t^\prime}\}.
\end{eqnarray}
It follows from (42) that the tensor $ \sigma _ {\alpha\beta}
({\bf r} ^ \prime, t ^ \prime | {\bf r}) $ does not depend on time
$ t $, if the potential energy $ V ({\bf r} _ 1\ldots {\bf r} _
N) $ from (8) does not depend on $ t $, what we mean. It is a
consequence of time uniformity. Therefore the designation is used
below
$$\sigma_{\alpha\beta}({\bf r}^\prime,
t^\prime|{\bf r})=\sigma_{\alpha\beta}({\bf r}^\prime,
t^\prime|{\bf r}, t).$$
Let us make a Fourier transform. Let us write down the electric
field as
\begin{equation}
\label{43}E_\alpha({\bf r}, t)=E_\alpha^{(+)}({\bf r},
t)+E_\alpha^{(-)}({\bf r}, t),
\end{equation}
where
\begin{equation}
\label{44}E_\alpha^{(+)}({\bf r}, t)={1\over(2\pi)^4}\int
d^3k\int_0^\infty d\omega E_\alpha({\bf k}, \omega)e^{i({\bf
k}{\bf r}-\omega t)},
\end{equation}
\begin{equation}
\label{45}E_\alpha^{(-)}({\bf r}, t)=(E_\alpha^{(+)}({\bf r},
t))^*,
\end{equation}
\begin{equation}
\label{46}E_\alpha({\bf k}, \omega)=\int d^3r\int_{-\infty}^\infty
dt E_\alpha({\bf r}, t)e^{i(-{\bf k}{\bf r}+\omega t)}.
\end{equation}
The splitting (43) is usually made to not use negative
 frequencies $ \omega $.

Let us introduce a Fourier-image $ \sigma _ {\alpha\beta} ({\bf
r} ^ \prime, t ^ \prime | {\bf r}) $ on variable $ {\bf r} ^
\prime, t ^ \prime $
\begin{equation}
\label{47}\sigma_{\alpha\beta}({\bf r}^\prime, t^\prime|{\bf r})=
\int d^3r^\prime\int_{-\infty}^\infty dt^\prime
\sigma_{\alpha\beta}({\bf r}^\prime, t^\prime|{\bf r}) e^{i(-{\bf
k}{\bf r}^\prime+\omega t^\prime)}.
\end{equation}
Then with the help of (41), (44) and (45) we obtain
\begin{eqnarray}
\label{48}\langle0| j_{1\alpha}({\bf r}, t)|0\rangle=
{1\over(2\pi)^4}\int d^3k\int_0^\infty d\omega\nonumber\\
\times\sigma_{\alpha\beta}({\bf k}, \omega|{\bf r})E_\beta({\bf
k}, \omega)e^{i{\bf k}{\bf r}-i\omega t} +c.c.
\end{eqnarray}
In a case of spatially - homogeneous systems, for example, bulk
semiconductor crystals
\begin{equation}
\label{49} \sigma_{\alpha\beta}({\bf k}, \omega) |{\bf
r})=\sigma_{\alpha\beta}({\bf k}, \omega).
 \end{equation}
 From (42), making
transformations (47) and executing integration on $ t ^ \prime, $
we obtain
\begin{eqnarray}
\label{50}&\sigma&_{\alpha\beta}({\bf k}, \omega|{\bf r}
)={ie^2\over\hbar\omega_gm_0^2}
\sum_{\eta}\nonumber\\
&\times&\{p_{cv\alpha}^*p_{cv\beta}F_{\eta}({\bf r}){\int
d^3r^\prime e^{-i{\bf k}{\bf r}^\prime}F_{\eta}^*({\bf r}-{\bf
r}^\prime)\over
\omega-\omega_\eta+i\gamma_\eta/2}\nonumber\\
&+&p_{cv\alpha}p_{cv\beta}^*F_{\eta}^*({\bf r}^\prime){\int
d^3r^\prime e^{-i{\bf k}{\bf r}^\prime}F_{\eta}({\bf r}-{\bf
r}^\prime)\over \omega+\omega_\eta+i\gamma_\eta/2}\}.
\end{eqnarray}
Let us notice that the conductivity tensor  has a property
\begin{equation}
\label{51} \sigma_{\alpha\beta}^*({\bf k}, \omega|{\bf
r})=\sigma_{\alpha\beta}(-{\bf k}, -\omega |{\bf r}).
 \end{equation}
Let us emphasize that the expressions (48) and (50) basically
allow to calculate average induced density of a current  at
monochromatic and at a pulse light irradiation and at any
direction of incident light, for example, not only at normal, but
also at slanting incidence of light on a QW's plane.

\section{The conductivity tensor in QW.}
It follows from (32) and (34) that for free EHPs in zero or
quantizing magnetic field the function $ F _ \eta ({\bf r}) $ may
be represented as a product
\begin{equation}
\label{52} F_\eta({\bf r})=Q_\pi({\bf r}_\perp)\phi_\chi(z),
 \end{equation}
 where $ \pi $ is the set of indexes
 $ v, {\bf k} _ {e\perp}, {\bf k} _ {v\perp} $ in $ H _ c = 0 $
 and $ v, n _ e, n _ v, k _ {ey}, k _ {vy} $ in quantizing magnetic field, $ \chi $ is the set
 of indexes $ v, l _ e, l _ v, $
\begin{equation}
\label{53}\phi_\chi(z)=\varphi_{l_e}^e(z)\varphi_{l_v}^v(z).
 \end{equation}
The splitting (52) is applicable also when the Coulomb
interaction of electrons and holes can essentially influence only
on movement of particles along $ z $ axis. It occurs under
condition of [19]
$$ a _ {exc} ^ 2> > a _ H ^ 2, $$
where
$$ a _ {exc} = {\hbar ^ 2\varepsilon _ 0\over \mu e ^ 2} $$
is the Wannier-Mott exciton radius  in absence of a magnetic
field, $ \varepsilon _ 0 $ is the static dielectric susceptibility
, $\mu=m_em_v/(m_e+m_v)$ is the effective mass, i.e. in a case  of
quantizing magnetic fields. At
$$ a_{exc} >> d, $$
i.e. for narrow bands the Coulomb forces poorly influence on
movement along an $ z $ axis and the functions $ \phi _ \chi (z)
$ look like (53). Otherwise
$$ a_ {exc} << d $$
the formula (53) is inapplicable. For GaAs
$$ a_ {exc} = 146 A, ~~~ a _ H ^ {res} = 57.2 A, $$
where $ a _ H ^ {res} $ corresponds to a magnetic field $ H_
{res}, $ at which has a place the magnetophonon resonance $
\Omega_{eH}=\omega_{LO}. $ Using (52) we obtain  from (50)
\begin{eqnarray}
\label{54}&\sigma&_{\alpha\beta}({\bf k}, \omega|{\bf
r})={ie^2\over\hbar\omega_gm_0^2}e^{-i{\bf k}{\bf r}}
\sum_{\eta}\nonumber\\
&\times&\phi_\chi(z)\int_{-\infty}^\infty
dz^\prime\phi_\chi(z^\prime)e^{ik_zz^\prime}\nonumber\\
&\times& \{p_{cv\alpha}^*p_{cv\beta}Q_\pi({\bf r}_\perp)\int d{\bf
r}_\perp^\prime Q_\pi^*({\bf r}_\perp^\prime){e^{i{\bf
r}_\perp{\bf
r}_\perp^\prime}\over \omega-\omega_\eta+i\gamma/2}\nonumber\\
&+&p_{cv\alpha}p_{cv\beta}^*Q_\pi^*({\bf r}_\perp)\int d{\bf
r}_\perp^\prime Q_\pi({\bf r}_\perp^\prime){e^{i{\bf r}_\perp{\bf
r}_\perp^\prime}\over \omega+\omega_\eta+i\gamma/2}\}.
\end{eqnarray}
For free EHPs at $ H _ c = 0 $
$$Q_\pi({\bf r}_\perp)={1\over S_0}e^{i({\bf k}_{e\perp} +{\bf
k}_{v\perp}){\bf r}_\perp}.$$
Substituting this expression in (54) and executing integration on
$ {\bf r} _ \perp ^ \prime $ we obtain
\begin{eqnarray}
\label{55}&\sigma&_{\alpha\beta}({\bf k}, \omega|{\bf
r})={ie^2\over\hbar\omega_gm_0^2S_0}e^{-ik_zz} \sum_{\eta}
\phi_\chi(z)R^*_\chi(z)
\nonumber\\
&\times&\left\{{p_{cv\alpha}^*p_{cv\beta}\delta_{{\cal K}_\perp,
k_\perp}\over\omega-\omega_\eta+i\gamma_\eta/2}+
{p_{cv\alpha}p_{cv\beta}^*\delta_{{\cal K}_\perp, -{\cal
k}_\perp}\over\omega+\omega_\eta+i\gamma_\eta/2}\right\},
\end{eqnarray}
where ${\cal K}_\perp=k_{e\perp}+k_{v\perp}$
\begin{equation}
\label{56}R_\chi(k_z)=\int_{-\infty}^\infty dz
e^{-ik_zz}\phi_\chi(z),
 \end{equation}
the energy $ \hbar\omega _ \eta $ is determined in (33).

For EHP in a quantizing magnetic field
\begin{eqnarray}
\label{57}Q_\pi({\bf
r}_\perp)&=&\Phi_{ne}(x+a_H^2k_{ey})\Phi_{nv}(x-a_H^2k_{vy})\nonumber\\
&\times&{1\over L_y}e^{i(k_{ey}+k_{vy})y}.
\end{eqnarray}
Let us substitute (57) in (54), integrate on variable $ y ^
\prime $ and make summation on indexes $ k _ {ey} $ and $ k _ {vy}
$, from which the energy $ \hbar\omega _ \eta $ (determined in
(35)) does not depend. It results in
\begin{eqnarray}
\label{58}&\sigma&_{\alpha\beta}({\bf k}, \omega|{\bf
r})={ie^2\over2\pi\hbar\omega_gm_0^2a_H^2}e^{-ik_zz}
\sum_{\xi}\phi_\chi(z)R^*_\chi(z)\nonumber\\
&\times&\left\{{p_{cv\alpha}^*p_{cv\beta}\Xi_{n_e, n_v}(k_x, k_y)
\over\omega-\omega_\xi+i\gamma_\xi/2}\right.\nonumber\\
 &+&\left.
{p_{cv\alpha}p_{cv\beta}^*\Xi_{n_e, n_v}(-k_x,
-k_y)\over\omega+\omega_\xi+i\gamma_\xi/2}\right\},
\end{eqnarray}
where
\begin{equation}
\label{59}\Xi_{n_e, n_v}(k_x, k_y)=\left|\int_{-\infty}^\infty
dt\Phi_{n_e}(t)\Phi_{n_v}(t-a_H^2k_y)e^{ik_xt}\right|^2,
\end{equation}
$ \xi $ is the set of indexes $ \chi, n _ e, n _ v, $ the energy $
\hbar\omega _ \xi = \hbar\omega _ \eta. $

Let us pay attention that $ \sigma _ {\alpha\beta} ({\bf k},
\omega | {\bf r}) $ in  QWs (at $ H_c = 0 $ or with $ {\bf H}_c$
directed along $ z $ axis) depends only on $ z $ what follows from
(55) and (58). It is caused by  that that the system is
inhomogeneous only along  $ z $ axis , perpendicular to a QW
plane. Thus,
\begin{equation}
\label{60}\sigma_{\alpha\beta}({\bf k}, \omega|{\bf
r})=\sigma_{\alpha\beta}({\bf k}, \omega|z).
 \end{equation}
The expression (48) with substitution in (55) or (58) is
applicable in case of incidence of light on QW under any angle to
 $ z $ axis for the monochromatic and for pulse irradiation.

\section{Normal light incidence on a QW surface.}

At a normal light incidence the electric field $ {\bf E} ({\bf r},
t) $ depends only on variables $ z $ and $ t $. Let us introduce
the Fourier-component of a field on variable $ t $
\begin{equation}
\label{61}E_\beta(z, \omega)=\int_{-\infty}^\infty dt e^{i\omega
t}E_\beta (z, t).
 \end{equation}
With the help of (48) and (55) it is possible to show that
average induced density of a current at $ H_c= 0 $ is equal
\begin{eqnarray}
\label{62}\langle0| j_{1\alpha}({\bf r}, t)|0\rangle= {1\over
2\pi}\left({e\over m_0}\right)^2{1\over \hbar\omega_g
S_0}\sum_\chi\phi_\chi(z)\nonumber\\
\times\int_{-\infty}^\infty d\omega e^{-i\omega t}
\int_{-\infty}^\infty dz^\prime\phi_\chi(z^\prime)E_\beta(z^\prime, \omega)\nonumber\\
\times\left\{p_{cv\alpha}^*p_{cv\beta}\sum_{{\bf
k}_\perp}(\omega-\omega_\kappa+i\gamma_\kappa/2)^{-1}\right.\nonumber\\
+\left. p_{cv\alpha}p_{cv\beta}^*\sum_{{\bf
k}_\perp}(\omega+\omega_\kappa+i\gamma_\kappa/2)^{-1}\right\},
\end{eqnarray}
where $ \kappa $ is the set of indexes $ \chi, {\bf k} _ \perp =
{\bf k} _ {e\perp} = - {\bf k} _ {v\perp} $,
\begin{equation}
\label{63}\omega_\kappa=\omega_g+\varepsilon_{l_e}^e/\hbar+\varepsilon_{l_v}^v/\hbar
+{\hbar k_\perp^2\over 2\mu}.
 \end{equation}
In (62) we passed from integration on $ \omega $ in limits from $
0 $ up to $ \infty $ to integration in limits from $ -\infty $ up
to $ \infty $, what is more convenient for concrete calculations,
because an integration contour may be closed in the top or bottom
half-plane.

In a  quantizing magnetic field with the help of (48) and (58) we
obtain
\begin{eqnarray}
\label{64}\langle0| j_{1\alpha}({\bf r}, t)|0\rangle= {1\over
(2\pi)^2}\left({e\over m_0}\right)^2{1\over \hbar\omega_g
a_H^2}\sum_\chi\phi_\chi(z)\nonumber\\
\times\int_{-\infty}^\infty d\omega e^{-i\omega t}
\int_{-\infty}^\infty dz^\prime\phi_\chi(z^\prime)E_\beta(z^\prime, \omega)\nonumber\\
\times\left\{p_{cv\alpha}^*p_{cv\beta}
\sum_n(\omega-\omega_\lambda+i\gamma_\lambda/2)^{-1}\right.\nonumber\\
+\left.
p_{cv\alpha}p_{cv\beta}^*\sum_n(\omega+\omega_\lambda+i\gamma_\lambda/2)^{-1}\right\},
\end{eqnarray}
where $ \lambda $ is the set of indexes $ \chi $ and $ n _ e = n
_ v = n $,
\begin{eqnarray}
\label{65}\omega_\lambda=\omega_g+\varepsilon_{l_e}^e/\hbar+\varepsilon_{l_v}^v/\hbar
+\Omega_{\mu H}(n+1/2),\nonumber\\
 \Omega_{\mu H}={|e|H_c\over\mu c}.
\end{eqnarray}
At a derivation of (64) the ratio was used
\begin{eqnarray}
\label{66}\Xi_{n_e, n_v}(k_x=0, k_y=0)=
\left|\int_{-\infty}^{\infty}dt\Phi_{n_e}(t)\Phi_{n_v}(t)\right|^2\nonumber\\
=\delta_{n_e,
n_v},
 \end{eqnarray}
which corresponds to the following selection rule: at a normal
incidence of light EHPs with identical Landau quantum numbers of
electrons and holes are excited.

At $ H _ c = 0 $ and a normal incidence of light EHPs with zero
quasi-momentum in a QW plane  are excited, therefore $ {\bf k} _
{e\perp} = - {\bf k} _ {h\perp} $, that follows from the
quasi-momentum conservation law in $ xy $ plane.

Let us notice that the expression (64) for quantizing magnetic
fields differs from (62) for $ H _ c = 0 $ only by replacement
 of the normalization area $ S _ 0 $ by $ 2\pi a _ H ^ 2 $ and
the index $ {\bf k} _ {\perp} $ by the index $ n $.

\section{The model simplifying expressions for an average current density.}

Further a model is used which was applied in [20-28].
 Vectors $ {\bf p} _ {cv} $ for two
degenerated valence bands $ v = I $ and $ v = II $ look like
\begin{eqnarray}
\label{67} {\bf p}_{cvI}={p_{cv}\over\sqrt{2}}({\bf e}_{x}- i{\bf
e}_{y}),\nonumber\\
{\bf p}_{cvII}={p_{cv}\over\sqrt{2} }({\bf e}_{x}+i{\bf e}_{y}),
\end{eqnarray}
where $ {\bf e} _ {x} $ and $ {\bf e} _ {y} $ are unite vectors
along axes $ x $ and $ y $, $ p _ {cv} $ is the real value. This
model corresponds to heavy holes in a semiconductor with the zinc
blend structure, if an axis is directed along an axis of symmetry
of the 4-th order [29,30]. If to use vectors of circular
polarization of stimulating light
\begin{eqnarray}
\label{68} {\bf e}_{l}={1\over\sqrt{2}}({\bf e}_{x}\pm i{\bf
e}_{x}),
\end{eqnarray}
the property of preservation of a polarization vector is performed
\begin{equation}
\label{69} \sum_{v=I, II}{\bf p}_{cv}^*({\bf e}_{l}{\bf
p}_{cv})=\sum_{v=I, II}{\bf p}_{cv}({\bf e}_{l}{\bf
p}_{cv}^*)={\bf e}_{l}p^2_{cv}.
\end{equation}
Wave functions $ \varphi _ {l _ v} ^ v $ and energy levels $
\varepsilon _ {l _ v} ^ v $ do not depend on numbers of valence
bands $ I $ and $ II $.

Using the model (67) and results (62) and (64) the expression for
average induced current density at $ H _ c = 0 $ in a quantizing
magnetic field we write down in an unified form
\begin{eqnarray}
\label{70}\langle0| j_{1\alpha}({\bf r}, t)|0\rangle= {ic\nu\over
8\pi^2}\sum_\rho\gamma_{r\rho}\phi_\rho(z)
\int_{-\infty}^\infty d\omega e^{-i\omega t}\nonumber\\
\times\int_{-\infty}^\infty dz^\prime
\phi_\rho(z^\prime)E_\alpha(z^\prime, \omega)\nonumber\\
\times\{(\omega-\omega_\rho+i\gamma_\rho/2)^{-1} +
(\omega+\omega_\rho+i\gamma_\rho/2)^{-1}\},
\end{eqnarray}
where $ \nu $ is the  refraction light factor and for $ H _ c = 0
$
\begin{equation}
\label{71} \gamma_{r\rho}=\gamma_r=4\pi{e^2\over\hbar
c\nu}{p_{cv}^2\over m_0^2}{1\over S_0\omega_g},
\end{equation}
$ \rho $ is the set of parameters $ l _ e, l _ h, {\bf k} _ \perp
$ and for quantizing magnetic fields
\begin{equation}
\label{72} \gamma_{r\rho}=\gamma_r=2{e^2\over\hbar
c\nu}{p_{cv}^2\over m_0^2}{1\over a_H^2\omega_g}=2{e^2\over\hbar
c\nu}{p_{cv}^2\over m_0^2}{\Omega_{0H}\over\hbar\omega_g},
\end{equation}
$ \Omega _ {0H} = | e | H/m _ 0c, ~~~\rho $ is the set of indexes
$ l _ e, l _ h, n $.

In the RHS  $ \gamma _ r $ is supplied with an index $ \rho $
though the RHSs of (71) and (72) do not depend on this index. It
is made for the expression (70) would be applicable and to another
situations, for example, to a case of the magnetopolaron
resonance in a quantizing magnetic field. The physical sense of $
\gamma _ {r\rho} $ will be opened below.

Let us notice that in the case of model (67) the important
property is performed
\begin{equation}
\label{73} div \langle0 | {\bf j} _ {1} ({\bf r}, t) | 0\rangle =
0,
\end{equation}
therefore the average induced charge density is equal 0 what
follows from the continuity equation.

\section{Calculation of a vector potential and an electric field.}

Knowing distribution of the average current density inside of QW,
it is possible to determine a vector potential according to the
known formula for retarded potentials (see, for example, [31, page
209])
\begin{equation}
\label{74} {\bf A}({\bf r}, t)={1\over c}\int d^3r^\prime {{\bf
j}({\bf r}^\prime, t-\nu|{\bf r}-{\bf r}^\prime|/c)\over|{\bf
r}-{\bf r}^\prime|}+{\bf A}_0({\bf r}, t).
\end{equation}
It follows from (70) that the dependence of a current density on
coordinates is determined only by  $ \phi _ \rho (z) $ in  the
sum on $ \varrho $.

Let us calculate the integral
\begin{equation}
\label{75} I_\rho(\omega, z)=\int d^3r^\prime
{\phi_\rho(z^\prime)e^{i\omega\nu|{\bf r}-{\bf
r}^\prime|/c}\over|{\bf r}-{\bf r}^\prime|},
\end{equation}
which is equal to
\begin{eqnarray}
\label{76} I_\rho(\omega, z)={2\pi
ic\over\omega\nu}\left\{\int_{-\infty}^z dz^\prime
\phi_\rho(z^\prime)e^{i\kappa(z-z^\prime)}\right.\nonumber\\
+\left.\int^{\infty}_z dz^\prime
\phi_\rho(z^\prime)e^{-i\kappa(z-z^\prime)}\right\},
\end{eqnarray}
where $ \kappa = \omega\nu/c $. Using (70), (74) and (76) we
obtain that the vector potential is equal
\begin{eqnarray}
\label{77} &A&_\alpha(z, t)=-{c\over
4\pi}\int_{-\infty}^\infty{d\omega\over\omega}e^{-i\omega
t}\nonumber\\
&\times&\sum_\rho\gamma_{r\rho}\int_{-\infty}^\infty
dz^\prime\varphi_\rho(z^\prime)E_\alpha(z^\prime,
\omega)\nonumber\\
&\times&\left\{\int_{-\infty}^z dz^\prime
\phi_\rho(z^\prime)e^{i\kappa(z-z^\prime)} +\int^{\infty}_z
dz^\prime
\phi_\rho(z^\prime)e^{-i\kappa(z-z^\prime)}\right\}\nonumber\\
&\times&\{(\omega-\omega_\rho+i\gamma_\rho/2)^{-1}+(\omega+\omega_\rho+i\gamma_\rho/2)^{-1}\}
\nonumber\\ &+&A_{0\alpha}(z, t).
\end{eqnarray}
It is specified above that an average charge  density in a case
of model (67) is equal 0, it means that the scalar potential $
\varphi $ is equal 0 also,  therefore
\begin{equation}
\label{78} {\bf E} (z, t) = - {1\over c} {\partial {\bf A} (z, t)
\over\partial t},
\end{equation}
and accordingly we obtain
\begin{eqnarray}
\label{79} &E&_\alpha(z, t)=-{i\over 4\pi}\int_{-\infty}^\infty
d\omega e^{-i\omega
t}\sum_\rho\gamma_{r\rho}\nonumber\\
&\times&\int_{-\infty}^\infty
dz^\prime\phi_\rho(z^\prime)E_\alpha(z^\prime,
\omega)\nonumber\\
&\times&\left\{\int_{-\infty}^z dz^\prime
\phi_\rho(z^\prime)e^{i\kappa(z-z^\prime)} +\int^{\infty}_z
dz^\prime
\phi_\rho(z^\prime)e^{-i\kappa(z-z^\prime)}\right\}\nonumber\\
&\times&\{(\omega-\omega_\rho+i\gamma_\rho/2)^{-1}+(\omega+\omega_\rho+i\gamma_\rho/2)^{-1}\}
\nonumber\\ &+&E_{0\alpha}(z, t),
\end{eqnarray}
where $ E _ {0\alpha} (z, t) $ is the exciting field. Let us make
the Fourier transform of the left and right parts of (79) using
(61). We have
\begin{eqnarray}
\label{80} &E&_\alpha(z, \omega)=-{i\over 2
}\sum_\rho\gamma_{r\rho}\int_{-\infty}^\infty
dz^\prime\phi_\rho(z^\prime)E_\alpha(z^\prime,
\omega)\nonumber\\
&\times&\left\{\int_{-\infty}^z dz^\prime
\phi_\rho(z^\prime)e^{i\kappa(z-z^\prime)} +\int^{\infty}_z
dz^\prime
\phi_\rho(z^\prime)e^{-i\kappa(z-z^\prime)}\right\}\nonumber\\
 &+&E_{0\alpha}(z, \omega).
\end{eqnarray}
Thus, we have obtained the integral equation for
Fourier-components of the electric field.

Let us write down the exciting field in the form
\begin{equation}
\label{81} {\bf E} _ 0 (z, t) = E _ 0 {\bf e} _ l\int _ {-\infty}
^ \infty D\omega e ^ {-i\omega p} D _ 0 (\omega) + c.c.,
\end{equation}
where $ p = t-\omega\nu z/c. $

For a monochromatic excitation with frequency $ \omega _ l $
$$ D _ 0 (\omega) = \delta (\omega-\omega _ l). $$
$ D _ 0 (\omega) $  also may correspond to pulses of any durations
and forms. It follows from (81)
\begin{equation}
\label{82} E _ {0\alpha} (z, \omega) = 2\pi E _ 0 e ^ {i\omega
\nu z/c } \{{\bf e} _ lD _ 0 (\omega) + {\bf e} _ l ^ *D _ 0
(-\omega) \}.
\end{equation}
Let us write down the required solution as
\begin{equation}
\label{83} {\bf E} (z, t) = {{\bf E} _ l\over2\pi} \int _
{-\infty} ^ \infty d\omega e ^ {-i\omega t} {\cal E} (z, \omega)
+ c.c..
\end{equation}
Then for $ {\cal E} (z, \omega) $ we obtain the equation
\begin{eqnarray}
\label{84}&{\cal E}&(z, \omega)=-{i\over 2
}\sum_\rho\gamma_{r\rho}\int_{-\infty}^\infty
dz^\prime\phi_\rho(z^\prime){\cal E}(z^\prime,
\omega)\nonumber\\
&\times&\left\{\int_{-\infty}^z dz^\prime
\phi_\rho(z^\prime)e^{i\kappa(z-z^\prime)} +\int^{\infty}_z
dz^\prime
\phi_\rho(z^\prime)e^{-i\kappa(z-z^\prime)}\right\}\nonumber\\
&\times&\{(\omega-\omega_\rho+i\gamma_\rho/2)^{-1}+
(\omega+\omega_\rho+i\gamma_\rho/2)^{-1}\}\nonumber\\
 &+&2\pi E_{0}e^{i\kappa z}D_0(\omega).
\end{eqnarray}

\section{The approximation of the infinitely deep QW.}

For the greater simplicity and demonstrativeness  of solutions
 we consider a case of an infinitely deep QW, when wave
functions $ \varphi _ l (z) $ of electrons and holes
 are strictly limited by limits of QW and there is no any their penetration
 in a barrier, i.e. for free EHPs
%

$$\varphi_l(z)= {2\over\sqrt{d}} \sin({lz\pi\over d}+{l\pi\over
2}), l=1, 2,\ldots; -{d\over2}\le z\le{d\over2}$$
 $$\varphi_l(z)=
0; z\le-{d\over 2}, z\ge{d\over2},$$
\begin{equation}
\label{85}  \varepsilon_l^e={\hbar^2\pi^2l^2\over 2m_ed^2},~~
\varepsilon_l^h={\hbar^2\pi^2l^2\over 2m_hd^2}.
\end{equation}
Then with the help of (84) we obtain
\begin{eqnarray}
\label{86}&{\cal E}&(z, \omega)=-{i\over 2
}\sum_\rho\gamma_{r\rho}\int_{-d/2}^{d/2}
dz^\prime\phi_\rho(z^\prime){\cal E}(z^\prime,
\omega)\nonumber\\
&\times&\left\{e^{i\kappa z}\int_{-d/2}^z dz^\prime
\phi_\rho(z^\prime)e^{-i\kappa z^\prime}\right.\nonumber\\
&+&\left.e^{-i\kappa z}\int^{d/2}_z dz^\prime
\phi_\rho(z^\prime)e^{i\kappa z^\prime}\right\}\nonumber\\
&\times&\{(\omega-\omega_\rho+i\gamma_\rho/2)^{-1}+
(\omega+\omega_\rho+i\gamma_\rho/2)^{-1}\}\nonumber\\
 &+&2\pi E_{0}e^{i\kappa z}D_0(\omega).
\end{eqnarray}

\section{The solution for a QW, which width is less than a light wave length.}

Let us consider the solution of (86) at
 $$ \kappa d << 1. $$
For monochromatic irradiation $ \kappa _ l = \omega _ l\nu/c $,
for pulse irradiation frequencies $ \omega $ are essential,
laying in an interval $ \pm\Delta\omega $ near the carrying pulse
frequency $ \omega _ l $. $ \Delta\omega $ is of the order of $
(\Delta t) ^ {-1} $, where $ (\Delta t) $ is the duration of a
light pulse. In any case $ \omega $ is of the order of $ \omega _
g $, where $ \hbar\omega _ g $ is the band gap of the
semiconductor. Let us search for the solution $ {\cal E} (z,
\omega) $ at the left and to the right of QW, where only plane
waves with frequencies $ \omega = c\kappa/\nu $ may exist. We
search for solutions of type
\begin{eqnarray}
\label{87}{\cal E}_{left}(z, \omega)&=&{\cal E}_{0}(z, \omega)+
\Delta{\cal E}_{left}(z, \omega),\nonumber\\
{\cal E}_{right}(z, \omega)&=&{\cal E}_{0}(z, \omega)+
\Delta{\cal E}_{right}(z, \omega),
\end{eqnarray}
\begin{eqnarray}
\label{88}\Delta{\cal E}_{left}(z, \omega)=2\pi E_0e^{-i\kappa
z}D(\omega),~~z\le -d/2,\nonumber\\
\Delta{\cal E}_{right}(z, \omega)=2\pi E_0e^{i\kappa
z}D(\omega),~~z\ge d/2.
\end{eqnarray}
We designate the field inside of QW  by indexes $ QW $
\begin{equation}
\label{89} {\cal E} _ {QW} (z, \omega) = {\cal E} _ {0} (z,
\omega) + \Delta {\cal E} _ {QW} (z, \omega).
\end{equation}
Let us determine the integral
$$\int _ {-d/2} ^ {d/2} dz\phi _ \rho (z) {\cal E} _ {QW} (z, \omega) $$
 included in the RHS of (86) in an approximation
$ \kappa d << 1 $. Substituting (89) in the integrand we obtain
\begin{eqnarray}
\label{90} \int_{-d/2}^{d/2} dz \phi_\rho(z){\cal E}_{QW}(z,
\omega)&=& \int^{d/2}_{-d/2} dz
\phi_\rho(z){\cal E}_{QW}(z, \omega)\nonumber\\
&+&2\pi E_{0}D_0(\omega)C_\rho,
\end{eqnarray}
where
$$C_\rho=\int^{d/2}_{-d/2} dz
\Phi_\rho(z).$$
The function $ \Delta {\cal E} _ {QW} (z) $ is unknown, but on
QW's borders we have
\begin{eqnarray}
\label{91}\Delta{\cal E}_{QW}(-d/2, \omega)&=&\Delta{\cal
E}_{left}(-d/2, \omega)=2\pi E_0e^{i\kappa
d/2}D(\omega),\nonumber\\
\Delta{\cal E}_{QW}(d/2, \omega)&=&\Delta{\cal E}_{right}(d/2,
\omega)\nonumber\\
&=&2\pi E_0e^{i\kappa d/2}D(\omega).
\end{eqnarray}
From (91) it is clear that at $ \kappa d << 1 $
\begin{equation}
\label{92}\Delta{\cal E}_{QW}(z, \omega)\simeq 2\pi E_0D( \omega).
\end{equation}
Substituting (92) in (90) we obtain
\begin{equation}
\label{93} \int^{d/2}_{-d/2} dz \phi_\rho(z){\cal E}_{QW}(z,
\omega)= 2\pi E_{0}(D_0(\omega)+D(\omega))'_\rho.
\end{equation}
Using the first equality from (88) we write down the equation
(86) for $ z < -d/2 $. In this area the integral
$$ \int ^ {z} _ {-d/2} dz ^ \prime e ^ {-i\kappa z ^ \prime
} \phi _ \rho (z ^ \prime), $$
included in the RHS of (86), is equal 0, and the integral
$$\int _ {z} ^ {d/2} dz ^ \prime e ^ {i\kappa z ^ \prime
} \phi _ \rho (z ^ \prime) = \int ^ {d/2} _ {-d/2} dz ^ \prime
\phi _ \rho (z ^ \prime) = C _ \rho. $$
We obtain the equation for $ D (\omega) $ the solution of which is
\begin{equation}
\label{94} D(\omega)=-{4\pi \chi(\omega)D_0(\omega)\over 1+4\pi
\chi(\omega)},
\end{equation}
\begin{eqnarray}
\label{95}\chi(\omega)&=&{i\over
8\pi}\sum_\rho\gamma_{r\rho}C_\rho^2\{(\omega-\omega_\rho+i\gamma_\rho/2)^{-1}\nonumber\\
&+&(\omega+\omega_\rho+i\gamma_\rho/2)^{-1}\}.
\end{eqnarray}
The equation (86) for $ z> d/2 $ also results in (95).

In a case of free movement of electrons and holes along $ z $
axis, when (53)is carried out,
$$C_\rho=\delta_{l_e, l_h}$$
and
$$\chi(\omega)={i\over
8\pi}\sum_{\rho_0}\gamma_{r\rho_0}\{(\omega-\omega_{\rho_0}+i\gamma_{\rho_0}/2)^{-1}$$
$$+(\omega+\omega_{\rho_0}+i\gamma_{\rho_0}/2)^{-1}\}.$$
 $ \rho _ 0 $ is the
set of indexes at $ l _ e = l _ h = l $, i.e. the set $ l, {\bf k}
_ \perp $ for $ H _ c = 0 $ and $ l, n $ for the case of a
quantizing magnetic fields.

The energy levels are accordingly equal
\begin{eqnarray}
\label{96}\omega_{\rho_0}=\bar{\omega}_{gl}+{\hbar^2k^2_\perp\over2\mu},
~~\omega_{\rho_0}=\bar{\omega}_{gl}+\Omega_\mu(n+1/2),
\end{eqnarray}
where
$$\bar {\omega} _ {gl} = \omega _ {g} + {\hbar ^ 2\pi ^ 2l ^ 2\over2\mu d ^ 2}. $$
For electric fields at the left and to the right of QW we obtain
the expressions
\begin{equation}
\label{97}{\bf E}_{left(right)}(z, t)={\bf E}_{0}(z, t)+
\Delta{\bf E}_{left(right)}(z, t),
\end{equation}
\begin{eqnarray}
\label{98}
 \Delta{\bf E}_{left(right)}(z,
t)&=& E_{0}{\bf e}_l\int_{-\infty}^\infty d\omega e^{-i\omega(t\pm
z\nu/c)}D(\omega)\nonumber\\&+&c.c.,
\end{eqnarray}
where the superscript concerns to an index $ left $, the subscript
- to an index $ right $. With the help of expressions (97) and
(98) it is possible to obtain formulas for reflected and absorbed
by QW light fluxes in case of any number of energy levels in QW,
any form of exciting pulse (including monochromatic irradiation)
and any ratio of parameters $ \gamma _ r $ and $ \gamma $ (
radiative and non-radiative broadenings of excitations). It
follows from (98) that the induced fields $ \Delta {\bf E} _
{left} (z, t) $ and $ \Delta {\bf E} _ {right} (z, t) $ differ
only by direction of movement.

\section{Solutions for QW, which width is comparable to a light wave length.
The first order on electron-light interaction.}

The electric field $ {\bf E} (z, t) $ may be spread out in a
series on electron-light interaction
\begin{equation}
\label{99}
 {\bf E}(z, t)={\bf E}_{0}(z, t)+{\bf E}_{1}(z, t)+{\bf E}_{2}(z, t) +...,
\end{equation}
where $ {\bf E} _ {0} (z, t) $ is the exciting field. It is
possible to obtain the following orders  by iterations, using the
equation (86).

In the first order we have
\begin{equation}
\label{100}
 {\bf E}_{1}(z, t)={{\bf e}_l\over2\pi}\int_{-\infty}^\infty d\omega e^{-i\omega t}
 {\cal E}_1(z, \omega)+c.c.,
\end{equation}
\begin{eqnarray}
\label{101}{\cal E}_1&(&z,
\omega)=-{i\over2}\sum_\rho\gamma_{r\rho}
\int_{-d/2}^{d/2}dz^\prime\phi_\rho(z^\prime){\cal E}_0(z^\prime,
\omega)\nonumber\\
&\times&\left\{e^{i\kappa z}\int_{-d/2}^{z}dz^\prime e^{-i\kappa
z^\prime}
\phi_\rho(z^\prime)\right.\nonumber\\&+&\left.e^{-i\kappa
z}\int_{z}^{d/2}dz^\prime e^{i\kappa z^\prime}
\phi_\rho(z^\prime)\right\}\nonumber\\
&\times&\{(\omega-\omega_\rho+i\gamma_\rho/2)^{-1}+(\omega+\omega_\rho+i\gamma_\rho/2)^{-1}\}.
\end{eqnarray}
Using definition
\begin{equation}
\label{102}
 {\cal E} _ {0} (z, \omega) = 2\pi E _ 0 e ^ {i\kappa z} D _ 0 (\omega),
\end{equation}
with the help of (100) we obtain
\begin{eqnarray}
\label{103} &E&_1(z, \omega)=-iE_0{\bf e}_l
\int_{-\infty}^{infty}d\omega e^{-i\omega
t}\nonumber\\
&\times&D_0(\omega)\sum_\rho(\gamma_{r\rho/2})R_\rho^*(\kappa)\nonumber\\
&\times&\left\{e^{i\kappa z}\int_{-d/2}^{z}dz^\prime e^{-i\kappa
z^\prime}
\phi_\rho(z^\prime)\right.\nonumber\\&+&\left.e^{-i\kappa
z}\int_{z}^{d/2}dz^\prime e^{i\kappa z^\prime}
\phi_\rho(z^\prime)\right\}\nonumber\\
&\times&\{(\omega-\omega_\rho+i\gamma_\rho/2)^{-1}
+(\omega-\omega_\rho+i\gamma_\rho/2)^{-1}\}\nonumber\\ &+&c.c.,
\end{eqnarray}
where the designation is introduced
\begin{equation}
\label{104} R _ \rho (\kappa) = \int _ {-d/2} ^ {d/2} dz\phi _
\rho (z) e ^ {-i\kappa z}.
\end{equation}
For free electrons and holes in an infinitely deep QW
\begin{equation}
\label{105} R _ \rho (\kappa) = \int _ {-d/2} ^ {d/2} dz\varphi _
{l _ e} (z) \varphi _ {l _ h} (z) e ^ {-i\kappa Z}.
\end{equation}
 The functions $ \varphi _ l (z) $ are determined in (85). From
(103) we obtain that the fields at the left and to the right of
QW are accordingly equal
\begin{eqnarray}
\label{106} &{\bf E}&_{1left}(z, t)=-iE_0{\bf e}_l
 \int_{-\infty}^{\infty}d\omega e^{-i\kappa z-i\omega
t}D_0(\omega)\nonumber\\
&\times&\sum_\rho(\gamma_{r\rho/2})(R_\rho^*(\kappa))^2\nonumber\\
&\times&\{(\omega-\omega_\rho+i\gamma_\rho/2)^{-1}
+(\omega+\omega_\rho+i\gamma_\rho/2)^{-1}\}\nonumber\\ &+&c.c.,
\end{eqnarray}
\begin{eqnarray}
\label{107}&{\bf E}&_{1right}(z, t)=-iE_0{\bf e}_l
 \int_{-\infty}^{\infty}d\omega e^{i\kappa z-i\omega
t}D_0(\omega)\nonumber\\
&\times&\sum_\rho(\gamma_{r\rho/2})|R_\rho^*(\kappa)|^2\nonumber\\
&\times&\{(\omega-\omega_\rho+i\gamma_\rho/2)^{-1}
+(\omega+\omega_\rho+i\gamma_\rho/2)^{-1}\}\nonumber\\ &+&c.c..
\end{eqnarray}
And for a field $ {\bf E} _ {1QW} (z, t) $ inside of QW it is
necessary to use (103). From (103) - (107) it follows that in a
case of wide QW there is allowable a creation of EHPs with
quantum numbers
\begin{equation}
\label{108} l _ e\ne l _ h,
\end{equation}
and also it appears there a dependence of fields on a  QW's width
$ d $ contained in factors $ R _ \rho (\kappa) $. It is possible
to show that, if to use functions (85), the factor $ R _ \rho ^
* (\kappa) /R _ \rho (\kappa) $, included in the
relation $ {\bf E} _ {1left} (z, t) / {\bf E} _ {1right} (z, t), $
depends on indexes $ l _ e $ and $ l _ h $ as follows: if $ l _ e
$ and $ l _ h $ are of identical parity, $ R _ \rho ^ * (\kappa)
/R _ \rho (\kappa) = 1 $, if $ l _ e $ and $ L _ h $ are of
different parity, $ R _ \rho ^ * (\kappa) /R _ \rho (\kappa) = -1
$.

Substituting the first order result (101) in the RHS of (86) one
obtains  the second order result, etc. So it is possible to
calculate all series (99). But we apply another method for
calculations of fields in case of wide QWs, i.e. under condition
of $ \kappa d\geq 1 $.

\section{Solutions for wide QW in a case of one energy level.}
In a case when one energy level is essential the equation (86)
may be solved exactly. Introducing designations
$$\omega _ \rho = \omega _ 0, ~~ \gamma _ \rho = \gamma, ~~
\phi _ \rho (z) = \phi (z), ~~ \gamma _ {r\rho} = \gamma _ r $$
 we rewrite (86) as
\begin{eqnarray}
\label{109}{\cal E}&(&z, \omega)=-{i\over2} \gamma_r
M(\omega)\left\{ e^{i\kappa z}\int_{-d/2}^{z}dz^\prime e^{-i\kappa
z^\prime}\phi(z^\prime)\right.\nonumber\\
&+&\left.e^{-i\kappa z}\int_{z}^{d/2}dz^\prime e^{i\kappa
z^\prime}\phi(z^\prime)\right\}\nonumber\\
&\times&\{(\omega-\omega_0+i\gamma/2)^{-1}+(\omega+\omega_0+i\gamma/2)^{-1}\}\nonumber\\
&+&2\pi E_0e^{i\kappa z}D_0(\omega),
\end{eqnarray}
where the designation is introduced
\begin{equation}
\label{110} M (\omega) = \int _ {-d/2} ^ {d/2} dz ^ \prime\phi (z
^ \prime) {\cal E} (z ^ \prime, \omega).
\end{equation}
Let us multiply both parts of (109) on $ \phi (z) $ and integrate
on $ z $ in limits from $ -d/2 $  up to $ d/2 $. It results in an
equation for $ M (\omega) $, which solution is
\begin{eqnarray}
\label{111}&M&(\omega)=2\pi E_0D_0(\omega)R^*(\kappa) \{1+{i\over
2}\gamma_rJ(\kappa)\nonumber\\
&\times&[(\omega-\omega_0+i\gamma/2)^{-1}\nonumber\\
&+&(\omega+\omega_0+i\gamma/2)^{-1}]\}^{-1},
\end{eqnarray}
where
\begin{eqnarray}
\label{112}J(\kappa)&=&\int_{-d/2}^{d/2}dz \phi(z)\left\{
e^{i\kappa z}\int_{-d/2}^{z}dz^\prime e^{-i\kappa
z^\prime}\phi(z^\prime)\right.\nonumber\\
&+&\left.e^{-i\kappa z}\int_{z}^{d/2}dz^\prime e^{i\kappa
z^\prime}\phi(z^\prime)\right\}.
\end{eqnarray}
It is possible to show that
\begin{equation}
\label{113} J (\kappa) = | R (\kappa) | ^ 2 + iQ (\kappa).
\end{equation}
Substituting (111) in (109) we obtain the solution of our task.
Using (83) we find induced electric fields at the left and to the
right of QW
\begin{eqnarray}
\label{114}&\Delta&{\bf E}_{left}(z, t)=-i{\bf e}_lE_0(\gamma_r/2)
\int_{-\infty}^\infty d\omega e^{-i\kappa z-i\omega
t}D_0(\omega)\nonumber\\
&\times&(R^*(\kappa))^2[(\omega-\omega_0+i\gamma/2)^{-1}
+(\omega+\omega_0+i\gamma/2)^{-1}]\nonumber\\
&\times&\{1+i(\gamma_r/2)(|R(\kappa)|^2+i Q(\kappa))\nonumber\\
&\times&[(\omega-\omega_0+i\gamma/2)^{-1}+(\omega+\omega_0+i\gamma/2)^{-1}]\}^{-1},
\end{eqnarray}
\begin{eqnarray}
\label{115}&\Delta&{\bf E}_{right}(z, t)=-i{\bf
e}_lE_0(\gamma_r/2) \int_{-\infty}^\infty d\omega e^{i\kappa
z-i\omega
t}D_0(\omega)\nonumber\\
&\times&(R^*(\kappa))^2[(\omega-\omega_0+i\gamma/2)^{-1}+(\omega+\omega_0+i\gamma/2)^{-1}]
\nonumber\\
&\times&\{1+i(\gamma_r/2)(|R(\kappa)|^2+i Q(\kappa))\nonumber\\
&\times&[(\omega-\omega_0+i\gamma/2)^{-1}+(\omega+\omega_0+i\gamma/2)^{-1}]\}^{-1}.
\end{eqnarray}
The value
\begin{equation}
\label{116} \tilde {\gamma} _ r (\omega) = \gamma _ r | R
(\kappa) | ^ 2
\end{equation}
at $ \omega = \omega _ 0 $ and at the account of (72) coincides
with calculated in [27]  radiative broadening of EHP in a
quantizing magnetic field at $ N _ e = n _ h = n, {\bf K} _ \perp
= 0 $ for any values $ \omega _ 0\nu d/c $.

Neglecting the non-resonant contribution $ (\omega + \omega _ 0 +
i\gamma/2) ^ {-1} $, we obtain from (114) and (115) the results
[26] \footnote {In [26] in the formulas (47) and (48) the typing
errors are admitted: instead of $ \tilde {\gamma} _ r $ it is
necessary to read $ \tilde {\gamma} _ re ^ {-i\kappa d/2}. $}
\begin{eqnarray}
\label{117}\Delta{\bf E}_{left}(z, t)=-i{\bf e}_lE_0
\int_{-\infty}^\infty d\omega\nonumber\\
\times {e^{-i\kappa z-i\omega
t}(\tilde{\gamma}_r(\omega)/2)D_0(\omega)e^{i\alpha}\over
\omega-(\omega_0+\Delta)+i(\gamma+\tilde{\gamma}_r(\omega))/2},
\end{eqnarray}
\begin{eqnarray}
\label{118}\Delta{\bf E}_{right}(z, t)=-i{\bf e}_lE_0
\int_{-\infty}^\infty d\omega\nonumber\\
\times {e^{i\kappa z-i\omega
t}(\tilde{\gamma}_r(\omega)/2)D_0(\omega)\over
\omega-(\omega_0+\Delta)+i(\gamma+\tilde{\gamma}_r(\omega))/2},
\end{eqnarray}
where
\begin{equation}
\label{119}e^{i\alpha}={R^*(\kappa)\over
R(\kappa)},~~\Delta=(\gamma/2)Q(\kappa).
\end{equation}
Let us notice that above this section we did not use the formula
(53), applicable only in case of free movement of electrons
 and holes along  $ z $ axis and assumed only
performance of (52).

In case of use (53) with substitution of functions (85) it is
possible to transform the expression for $ R (\kappa) $ and $ R ^
* (\kappa) $ to
$$R(\kappa)={1\over 2}\int_{-d/2}^{d/2}dz e^{-i\kappa z}$$
$$\times\{\cos[(\pi/d)(l_e-l_h)z+(\pi/2)(l_e-l_h)z]$$
$$-\cos[(\pi/d)(l_e+l_h)z+(\pi/2)(l_e+l_h)]\},$$
$$R^*(\kappa)={1\over 2}\int_{-d/2}^{d/2}e^{-i\kappa z}$$
$$\times\{\cos[(\pi/d)(l_e-l_h)z-(\pi/2)(l_e-l_h)]$$
$$-\cos[(\pi/d)(l_e+l_h)z-(\pi/2)(l_e+l_h)]\}.$$
In case of narrow QWs at $ \kappa d << 1 $ we have
$$ R (\kappa) = R ^ * (\kappa) = \delta _ {l _ e, l _ h}. $$
It means that light creates only EHPs with identical numbers of
the size quantization of electrons and holes (in a limit of
infinite deep QWs). In a case $ \kappa d\geq 1 $ EHPs are born
with various $ l _ e $ and $ l _ h $, i.e. there may be an
excitation of much more energy levels.

\section{Monochromatic and pulse excitation.}

In [7-15] the following expression is used for a light pulse
\begin{eqnarray}
\label{120}{\bf E}(z, t)=E_0({\bf e}_le^{-i\omega_lp}+{\bf
e}_l^*e^{i\omega_lp})\nonumber\\
\times\{\Theta(p)e^{-\gamma_{l1}p/2}+[1-\Theta(p)]e^{\gamma_{l2}p/2}\},
\end{eqnarray}
where $ \omega _ l $ is the carrying frequency, $ p = t-z\nu/c,
\Theta (p) $ is the Haeviside  function. Decomposing a pulse on
frequencies, we have
\begin{equation}
\label{121}{\bf E}(z, t)=E_0{\bf e}_l\int_{-\infty}^\infty d\omega
e^{-i\omega p}D_0(\omega)+c.c.,
\end{equation}
where
\begin{equation}
\label{122}D_0(\omega)={i\over
2\pi}[(\omega-\omega_l+i\gamma_{l1}/2)^{-1}-(\omega-\omega_l-i\gamma_{l1}/2)^{-1}].
\end{equation}
Under condition of $ \gamma _ {l1} = \gamma _ {l2} = \gamma _ {l}
$ the pulse is symmetric, its duration is of order $ \gamma _ l ^
{-1} $. At $ \gamma_l\rightarrow 0 $ we obtain
\begin{equation}
\label{123} D _ 0 (\omega) = \delta (\omega-\omega _ l),
\end{equation}
what corresponds to monochromatic irradiation. At $ \gamma _ {l2}
\rightarrow 0 $ the pulse is asymmetrical and has a very much
abrupt front.

The case of an monochromatic irradiation is considered in [26,27],
only asymmetric pulse - in [20-22], only symmetric pulse - in
[23,24,28], symmetric and asymmetric pulses - in [25].

\section{Conclusions.}
It is possible to allocate two most important results. First, the
expressions (48) and (50) for the averaged density of the induced
current is applicable to any semiconductor objects in case of any
number of energy levels of electronic excitations and at any form
of a stimulating light pulse and also at anyone direction of light
concerning crystal axes.

Second, (84) is the integral equation  for  Fourier-components of
the electric field in a case of normal incidence of light on a QW,
 which width can be comparable to the light wave length,
and number of energy levels of excitations is anyone, what in
particular corresponds to a QW in a quantizing magnetic field.
The equation is applicable both for monochromatic and for pulse
light excitation. With the help of these results it is possible
to solve a plenty of tasks on optics of low-dimensional
semiconductor objects.

\end{document}